\documentclass[pra,aps,showpacs,twocolumn,floatfix]{revtex4}
\usepackage{graphicx}
\usepackage[ansinew]{inputenc}
\usepackage{array}
\usepackage{color}
\usepackage{amsmath}
\usepackage{amsxtra}
\usepackage{amstext}
\usepackage{amssymb}
\usepackage{latexsym}
\usepackage{dsfont}
\usepackage{verbatim}

\begin{document}

\title{Laser phase noise effects on the dynamics of optomechanical resonators}
\author{Gregory A. Phelps}
\author{Pierre Meystre}
\affiliation{B2 Institute, Department of Physics and College of Optical Sciences\\The University of Arizona, Tucson, Arizona, 85721}
\date{\today}

\begin{abstract}
We investigate theoretically the influence of laser phase noise on the cooling and heating of a generic cavity optomechanical system. We derive the back-action damping and heating rates and the mechanical frequency shift of the radiation pressure-driven oscillating mirror, and derive the minimum phonon occupation number for small laser linewidths. We find that in practice laser phase noise does not pose serious limitations to ground state cooling. We then consider the effects of laser phase noise in a parametric cavity driving scheme that minimizes the back-action heating of one of the quadratures of the mechanical oscillator motion. Laser linewidths narrow compared to the decay rate of the cavity field will not pose any problems in an experimental setting, but broader linewidths limit the practicality of this back-action evasion method.
\end{abstract}


\maketitle

\section{Introduction}

The emerging field of cavity optomechanics~\cite{KippenbergVahala2007} is witnessing rapid and remarkable  progress, culminating recently in the cooling of micromechanical cantilevers to the ground state of motion~\cite{OConnell2010}. With the prospect of a broad variety of systems reaching that milestone in the near future, the emphasis of much current research is now shifting to ``beyond ground state" physics. Because cavity optomechanics is largely driven by the double goal of developing force sensors of extreme sensitivity and to investigate quantum effects in nanoscale (or larger) systems, a major near-term goal of that program involves the manipulation and control of the quantum state of these systems. Examples of particular interest include the preparation of quantum states, such as squeezed states, that allow us to circumvent the standard quantum limit, the generation of non-classical, macroscopically occupied phononic fields such as Fock states with large occupation number, and the realization of macroscopic quantum superpositions \cite{Armour2002, Marshall2003}. Quantum entanglement between two or more mechanical oscillators, or between mechanical oscillators and optical fields, is another goal with much promise for quantum metrology \cite{Helmerson2008}. In all of these situations, dissipation and decoherence are, of course, major obstacles that need to be understood and brought under control.

\begin{figure}[t]
\includegraphics[width= 0.4 \textwidth]{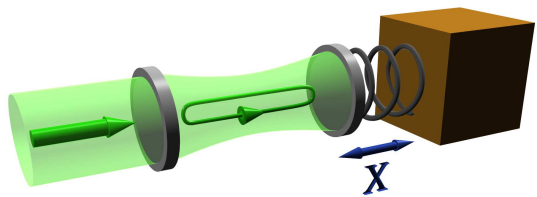}
\caption{(Color Online) Generic cavity optomechanical system.  The cavity consists of a highly reflective fixed input mirror and a small movable end mirror harmonically coupled to a support that acts as a thermal reservoir.  }
\end{figure}

Most optomechanical systems are comprised of a mechanical oscillator attached to a support that is either at room temperature or in a cryostat environment. In such systems, clamping losses are usually the dominant source of dissipation and decoherence, and major efforts are underway to control and minimize these losses. One approach that is currently receiving much attention is the use of ``all-optical" optomechanical systems comprised for instance of optically levitating micro-mirrors~\cite{Swati2010} or of dielectric micro-spheres~\cite{RomeroIsart2010, Chang2010}. The remarkable isolation, extremely long mechanical coherence times, high sensitivity to forces and displacements, as well as the ability to generate non-classical light and phononic fields in such systems are particularly promising features of these systems. In such situations, though, laser fluctuations, which are otherwise a minor concern when compared to clamping noise, become a major issue, perhaps `the' major issue.

The effect of laser phase noise in the cooling and coherent evolution of optomechanical systems has recently been studied in much detail by Rabl {\it et al.}~\cite{Rabl2009}, who concluded that while laser noise does pose a challenge to ground state cooling and the coherent transfer of single excitations between the optical cavity and the mechanical resonator, it is not a stringent limitation, in contrast to earlier predictions~\cite{Diosi2008}, see also Ref.~\cite{Yin2009}. The present paper expands on these results to consider not just the cooling regime, but also the regime of parametric instability --or more precisely mechanical amplification and regenerative
oscillations~\cite{instability} -- that can be reached for laser fields blue-detuned from the cavity resonance. It is known that this instability can lead to self-sustained oscillations and phononic lasing \cite{Vahala2009,Grudinin2010,Braginsky1980}. As such, this regime is particularly promising for the ``beyond ground state'' program, as it may result, when combined with phononic analogs of cavity QED, to the generation of non-classical phononic fields. We also consider the effects of laser noise on the parametric driving of the oscillator, a situation that may lead to back-action evading measurements of one quadrature of mechanical motion, and the possibility of generating a squeezed state of motion~\cite{Clerk2008}.

The paper is organized as follows: Section II introduces our model and establishes the notation. Section III discusses the effects of laser phase noise on back-action cooling and the optical spring effect, within a classical description of both the mirror motion and the intracavity light field. It then turns to a quantum description of the mirror motion to evaluate the minimum mean phonon number in the red-detuned driving regime. It also comments on  the unstable blue-detuned regime. Section IV discusses the parametric driving of the mechanical oscillator and evaluates the influence of a finite laser linewidth on the heating of the out-of-phase quadrature. Finally Section V is a summary and conclusion.

\section{Model system}

We consider a generic cavity optomechanical system modeled as a Fabry-P{\'e}rot cavity with one fixed input mirror and a harmonically bound movable end mirror connected to a support, see Fig.~1.  An incident laser beam of carrier frequency $\omega_\ell$, classical field amplitude $E(t)$ and power $P$ provides the desired radiation pressure to achieve cooling, an instability, or squeezing of the center-of-mass motion of the mirror. At the simplest level we describe the optical field inside the Fabry-P{\' e}rot as a single-mode field, coupled to the center-of-mass (COM) mode of motion of the moving mirror of oscillating frequency $\Omega$ and effective mass $M$ by the usual cavity optomechanical coupling. This system is described by the Hamiltonian~\cite{Law1995}
\begin{eqnarray}
\hat{H}&=& \hbar\Omega \hat{a}^\dagger \hat{a} +\hbar\omega_c \hat{b}^\dagger \hat{b} - \hbar g_0 \left(\hat{b}^\dagger \hat{b}-\langle\hat{b}^\dagger \hat{b}\rangle\right) \left(\hat{a}^\dagger+\hat{a}\right) \nonumber \\
&+&i\hbar\left[\hat{\eta}^\dagger(t) \hat{b}-\hat{b}^\dagger\hat{\eta}(t)\right]
 + \hat{H}_{\Gamma}+\hat{H}_{\kappa}
\label{H}
\end{eqnarray}
where $\hat{H}_{\Gamma}$ and $\hat{H}_{\kappa}$ describe the coupling of the mirrorCOM mode and the cavity field to reservoirs and account for dissipation at rates $\Gamma$ and $\kappa$, respectively. The bosonic creation and annihilation operators $\hat{a}^\dagger$ and $\hat{a}$ describe the COM phononic mode and $\hat{b}^\dagger$ and $\hat{b}$ describe the cavity field mode of frequency $\omega_c$. The optomechanical coupling coefficient is $g_0 = (\omega_c/L) x_{\rm zpt}$, where $x_{\rm zpt}= [\hbar/2M\Omega]^{1/2}$ is the ground state position uncertainty of the mechanical oscillator and $L$ is the equilibrium length of the Fabry-P{\'e}rot.

The optical driving rate $\hat{\eta}(t)$ of the intracavity field is given by~\cite{Giovannetti2001}
\begin{equation}
\hat{\eta}(t)= \sqrt{\frac{c \epsilon_0 \sigma \kappa}{\hbar \omega_\ell}} E(t) e^{-i \omega_\ell t+i\phi(t)}+\sqrt{\kappa} \hat{d}_{\rm in}(t)e^{-i\omega_\ell t},
\end{equation}
where $\sigma$ is the area of the incident beam and $\kappa$ the intrinsic cavity loss rate. Laser phase noise can be accounted for by a random phase $\phi(t)$ characterized in the case of a Lorentzian linewidth by the two-time correlation function
\begin{equation}
\label{noise}
\langle\dot{\phi}(t)\rangle_{\rm av} =  0, \mbox{        } \langle\dot{\phi}(t)\dot{\phi}(s)\rangle_{\rm av} = \sqrt{2 \gamma} \delta(t-s),
\end{equation}
where $\langle \rangle_{\rm av}$ denotes the classical ensemble average.  The bosonic noise operator $\hat{d}_{\rm in}(t)$, which accounts for quantum fluctuations of the classical laser field, satisfies the two-time correlations functions
\begin{equation}
\langle\hat{d}_{\rm in}^\dagger (t) \hat{d}_{\rm in}(s)\rangle = 0, \mbox{      } \langle\hat{d}_{\rm in} (t) \hat{d}_{\rm in}^\dagger(s)\rangle = \delta(t-s).
\label{corr_din}
\end{equation}
From Eq.~(\ref{H}) one readily obtains the Langevin equations of motion for the cavity field ($\hat{b} \rightarrow \hat{b}e^{i\omega_\ell t}$) and COM operators
\begin{eqnarray}
\dot{\hat{a}} &=& \left[- i \Omega-\frac{\Gamma}{2}\right] \hat{a}-\sqrt{\Gamma}\hat{a}_{\rm in}(t)+i g_{\rm 0} \left(\hat{b}^\dagger \hat{b} -\langle \hat{b}^\dagger \hat{b}\rangle \right) \\
\dot{\hat{b}} &=&\left[ i\Delta-\frac{\kappa}{2}\right] \hat{b}-\hat{\eta}(t)e^{i \omega_\ell t}+i g_{\rm 0} \hat{b} \left(\hat{a}^\dagger+\hat{a}\right),
\end{eqnarray}
where $\Delta = \omega_\ell-\omega_c$ is the detuning from cavity resonance.  From standard input-output formalism \cite{Walls1994}, the thermal input term $\hat{a}_{\rm in}(t)$ obeys the two-point correlations
\begin{eqnarray}
\langle\hat{a}_{\rm in}^\dagger (t) \hat{a}_{\rm in}(s)\rangle &=& n_M \delta(t-s) \nonumber \\
\langle\hat{a}_{\rm in}(t) \hat{a}_{\rm in}^\dagger(s)\rangle &=& (n_M+1)\delta(t-s),
\label{corr_ain}
\end{eqnarray}
where $n_M = k_b T_{\rm eff}/\hbar \Omega$ is the thermal occupation number of an oscillator of mechanical frequency $\Omega$ coupled to a thermal reservoir at temperature $T_{\rm eff}$.  With $\hat{b} = \bar{b}+\hat{d}$, where $\bar{b}$ is the classical part of the cavity field and $|\bar{b}|^2 \gg \langle \hat{d}^\dagger \hat{d} \rangle$, we easily obtain the linearized Langevin equations of motion
\begin{eqnarray}
\dot{\hat{a}} &\approx& \left[- i \Omega-\frac{\Gamma}{2}\right] \hat{a}-\sqrt{\Gamma}\hat{a}_{\rm in}+i g_{\rm 0} \left(\bar{b} \hat{d}^\dagger+\bar{b}^* \hat{d} \right) \label{LinearA} \\
\dot{\hat{d}} &\approx& \left[ i\Delta-\frac{\kappa}{2}\right] \hat{d}-\sqrt{\kappa}\hat{d}_{\rm in}(t)+i g_{\rm 0} \bar{b} \left(\hat{a}^\dagger+\hat{a}\right),
\label{LinearD}
\end{eqnarray}
where the classical amplitude $\bar{b}$ obeys the equation of motion
\begin{equation}
\dot{\bar{b}} = \left[i \Delta -\frac{\kappa}{2}\right] \bar{b}-\sqrt{\frac{c \epsilon_0 \sigma \kappa}{\hbar \omega_\ell}} E(t) e^{i\phi(t)}.
\label{LinearB}
\end{equation}

\section{Single frequency driving}

\subsection{Back-action cooling and optical spring effect}

It is well known that a driving laser tuned to the red side of the cavity resonance results in an increase in cavity damping and a concomitant cooling of the COM motion, while at the same time reducing the mirror oscillator frequency~\cite{KippenbergVahala2007}, the optical spring effect. To derive the radiation-pressure induced corrections to the damping and frequency shift in the presence of a finite laser linewidth we introduce the COM position and momentum operators in the familiar way as
\begin{eqnarray}
\hat{x} &=& x_{\rm zpt}(\hat{a}+\hat{a}^\dagger),\nonumber \\
\hat p&=&[i\hbar/(2 x_{\rm zpt})](\hat a^\dagger - \hat a )
\end{eqnarray}
and the scaled field mode operator
\begin{equation}
\hat \beta = \sqrt{\hbar \omega_c} \hat b.
\end{equation}

We consider first the situation where the mirror motion and the intracavity light field can both be described classically, $\hat{x} \rightarrow x$, $\hat p \rightarrow p$, $\hat \beta \rightarrow \beta$. With Eqs.~(5) and (6), the Langevin equations of motion describing the mirror motion and the intracavity field are then
\begin{eqnarray}
\label{classical_langevin}
&&\ddot{x}+\Gamma \dot{x}+\Omega^2 x = \frac{\left|\beta\left(t\right)\right|^2}{M L}-\frac{\left|\beta_0\right|^2}{M L}+\sqrt{\frac{2 k_b T_{\rm eff} \Gamma}{M}} \nu\left(t\right), \nonumber \\
&& \dot{\beta} = \left[i [\Delta+g_0 (x/x_{\rm zpt})]-\frac{1}{2}\kappa\right]\beta+\sqrt{\kappa P} e^{i \phi\left(t\right)},
\end{eqnarray}
where $P= c \epsilon_0 \sigma |E|^2$ is the driving laser power, $\nu(t)$ is a Gaussian noise process of zero mean, $\langle \nu(t)\nu(s)\rangle_{\rm av} = \delta(t-s)$, and $|\beta_0|^2$ is the mean intracavity field energy, given by
\begin{equation}
\left |\beta_0\right |^2 = P \frac{4(2\gamma+\kappa)}{(2\gamma+\kappa)^2+4\Delta^2}.
\label{beta0}
\end{equation}
For a finite laser linewidth, $\gamma$, the optical damping coefficient becomes
\begin{equation}
\label{Gamma opt}
\Gamma_{\rm opt} = P \left(\frac{\omega_c\kappa}{\Omega M L^2}\right) \frac{8\left[A_--A_+\right]}{\left[\left(2\gamma+\kappa\right)^2+4\Delta^2\right]},
\end{equation}
where we have assumed $\Gamma+\Gamma_{\rm opt} \ll \kappa$. Here,

\begin{figure}[t]
\includegraphics[width= 0.45 \textwidth]{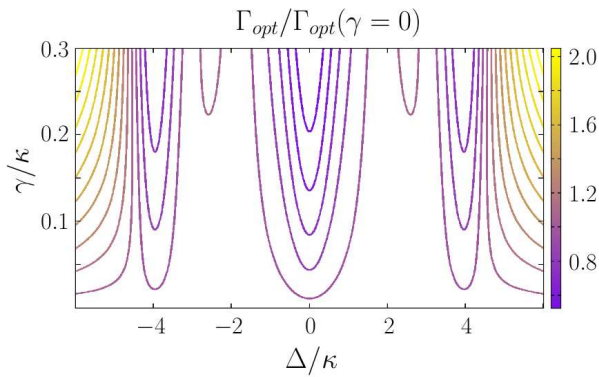}
   \caption{(Color Online) Radiation pressure induced damping as a function of laser linewidth $\gamma$ and detuning $\Delta$, normalized to the ideal case $\gamma = 0$.  We observe a range of linewidths that result in an {\em increase} in damping. This is a combined result of an increase in intracavity mean photon number and a broadening of the anti-stokes sideband, see text. In this example $\Omega = 4\kappa$. All rates are normalized to $\kappa$.}
\end{figure}

\begin{equation}
\label{A}
A_{\pm} = \frac{\left(\gamma +\kappa\right)\left(2\gamma +\kappa\right)^2
+2\gamma \left(\left(\Delta\mp\Omega\right)^2
+\Delta^2\right)+\kappa\Omega^2}{\left[\left(2\gamma +\kappa\right)^2
+4\left(\Delta\mp\Omega\right)^2\right]\left(\kappa^2+\Omega^2\right)}.
\end{equation}
The details of the derivation are given in Appendix A. The expression~(\ref{Gamma opt}) reduces to the results of Ref.~\cite{KippenbergVahala2007} for $\gamma \rightarrow 0$, as it should.

Similarly, the optically induced shift in the mirror COM frequency becomes
\begin{equation}
\label{omega}
\Delta \Omega_{\rm opt} = -P \left(\frac{\omega_{c}\kappa}{\Omega^2 M L^2}\right) \frac{2\left[B_+-B_-\right]}{\left[\left(2\gamma+\kappa\right)^2
+4\Delta^2\right]\left(\kappa^2+\Omega^2\right)},
\end{equation}
where
\begin{widetext}
\begin{equation}
\label{B}
B_\pm = \frac{\kappa\left(2\gamma+\kappa\right)^3+\kappa^2\left(2\Delta\pm\Omega\right)^2
+\left(8\gamma\Delta\kappa+4\Delta\Omega^2\right)\left(\Delta\pm\Omega\right)
-4\gamma^2\Omega^2}{\left(2\gamma+\kappa\right)^2+4\left(\Delta\pm\Omega\right)^2}.
\end{equation}
\end{widetext}

It is well known that for $\gamma =0$ that back-action damping is optimized for a laser red-detuned from the cavity resonance by the COM oscillation frequency $\Delta = -\Omega$. As would be intuitively expected a finite laser linewidth decreases $\Gamma_{\rm opt}$ for this optimal detuning. Somewhat surprisingly, though, an increase in laser linewidth can also result in an {\em increased} cooling for a small range of detunings $\Delta \neq -\Omega$  see Fig.~2.

One can gain an intuitive feeling for this unexpected behavior by first considering the coefficients $A_\pm(\Delta,\gamma)$ and recalling how they contribute to either cold damping or to a possible instability. Figure~3 shows $A_-(\Delta,\gamma)$ as a function of $\Delta$ for increasing values of $\gamma$. It is always positive, but its peak value, at $\Delta = -4\kappa$ for the parameters of the figure, decreases with increasing $\gamma$.  Since $A_+(\Delta,\gamma) = A_-(-\Delta,\gamma)$, $A_+(\Delta,\gamma)$ has the same behavior for $\Delta \rightarrow -\Delta$. For a red-detuned driving laser at the peak detuning $\Delta=-4 \kappa$ the $A_-$ contribution to Eq.~(\ref{Gamma opt}) dominates over the $A_+$ contribution, leading to an increase in the damping rate of the mirror and in cooling. For a a blue-detuned laser, on the other hand, the $A_+$ contribution dominates, leading to decreased mirror damping and to the onset of an instability for appropriate parameters.
\begin{figure}[t]
\center
\includegraphics[width= 0.44 \textwidth]{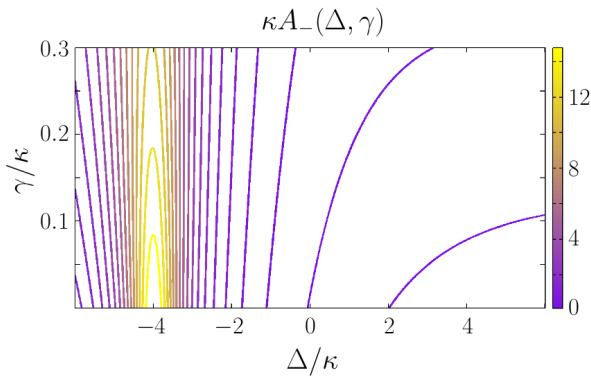}
\label{aplus}
\caption{(Color Online) $A_-(\Delta, \gamma)$ as a function of the laser-Fabry-P{\'e}rot detuning and the laser linewidth for $\Omega = -4 \kappa$. All rates are in dimensionless units.}
\end{figure}

The complex behavior of back-action damping as a function of the laser linewidth $\gamma$ can then be understood as a result of a delicate balance between the dependence of $A_\pm(\Delta,\gamma)$ on $\Delta$ and the dependence of the intensity of the relevant spectral components of the intracavity field on $\gamma$ (see Appendix A). Several examples of this dependence are shown in Fig.~4.
\begin{figure}[t]
\center
\includegraphics[width=0.45 \textwidth]{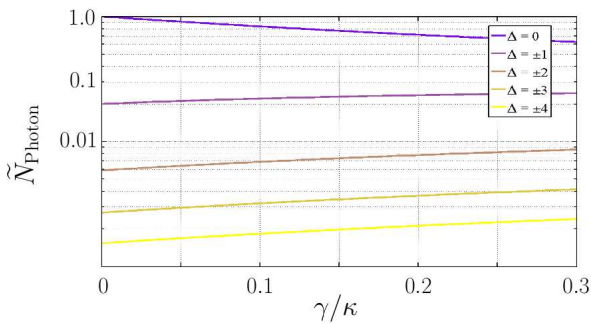}
\caption{(Color Online) Normalized intracavity intensity ($\widetilde{N}_{\rm Photon} = (\kappa/4 P) |\beta_0|^2$) as a function of the laser linewidth $\gamma$ for several values of the detuning $\Delta$. All rates are normalized to $\kappa$.}
\end{figure}
For large detunings, $|\beta_0(\Delta, \gamma)|^2$ increases with $\gamma$, but this increase is not quite linear, and, of course, neither is the dependence of $A_\pm(\Delta,\gamma)$ on $\Delta$. A finite laser linewidth will therefore result in increased back-action damping for
\begin{eqnarray}
G(\Delta,\gamma) &>& G(\Delta,0),  \\
G(\Delta,\gamma) &=& \left|\int_{- \infty}^{\infty} (A_-(\nu)-A_+(\nu)) |\beta_0(\nu-\Delta, \gamma)|^2 d\nu\right|. \nonumber
\end{eqnarray}
For the reversed inequality, the laser linewidth results in a decrease in back-action damping. The situation at resonance is slightly different. Here the decrease in cold damping is simply a result in the decrease in the intensity of the spectral components about $\Delta =0$ for increasing $\gamma$, see Fig.~4.

\begin{figure}[t]
\includegraphics[width= 0.45 \textwidth]{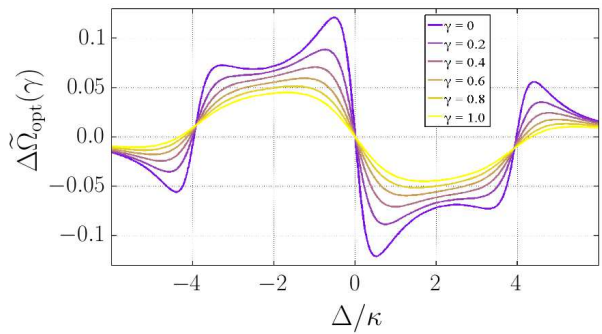}
   \caption{(Color Online) Radiation pressure induced mechanical frequency shift as a function of the detuning $\Delta$ and laser linewidth $\gamma$. Here $\Delta\widetilde{\Omega}_{\rm opt} = \Delta\Omega_{\rm opt}/\left(2 P \omega_c / \Omega^2 M L^2 \kappa\right)$, and all frequencies are normalized to $\kappa$.  Note the shift in the ``zero" mechanical frequency shift near $\Delta = \pm 4 \kappa$.}
\end{figure}

Similar arguments can be invoked to understand the behavior of the mechanical frequency shift. We observe an increase in $\Delta\Omega_{\rm opt}$ for $\Delta < -\Omega$, but a decrease for $-\Omega < \Delta < 0$  for a finite laser linewidth.  This is illustrated in Fig.~5, which shows the radiation pressure induced mechanical frequency shift in the presence of laser linewidth as a function of relative detuning, $\Delta/\kappa$, and for various laser linewidths, $\gamma/\kappa$.  For most of the relevant parameter range we have $|\Delta\Omega_{\rm opt}(\gamma)| <  |\Delta\Omega_{\rm opt}(\gamma = 0)|$, a result of the increase in the ``effective'' cavity linewidth from $\kappa$ to $\kappa + 2 \gamma$ due to the finite laser linewidth.  This increase is equivalent to the softening of the radiation pressure-induced potential.  In the good cavity limit $\Omega \gg \kappa$, and for $\Delta = -\Omega$, there is always an increase in the mechanical frequency shift.  Specifically, for small laser linewidths ($\gamma \ll \kappa$) the mechanical frequency shift increase is given by
\begin{equation}
\Delta\Omega_{\rm opt}(\gamma,\Delta = -\Omega) \approx \Delta\Omega_{\rm opt}(0,\Delta=-\Omega)+\left(\frac{2 P \omega_{c}\kappa}{\Omega^2 M L^2}\right) \frac{\gamma\kappa}{\Omega^2}.
\end{equation}
The increase in $\Delta\Omega_{\rm opt}$ has its origin in a change in the position of its zero for negative detunings, again see Fig.~5.

Similar effects occur on the heating side ($\Delta > 0$).  The heating rate of the mirror is reduced near $\Delta = \Omega$ for finite laser linewidth, see Fig~2, but in analogy to the situation on the cooling side, we note an increase in the heating rate for a range of detunings $\Delta > \Omega$.  As expected, we also observe a decrease in the mechanical frequency shift for finite laser linewidths. This indicates that the detuning required to maximize the optical spring effect depends on both the mechanical frequency $\Omega$ and on the linewidth of the input laser. These considerations may play a role in the optimization of the operation of optomechanical phonon lasers.

\subsection{Minimum phonon occupation number}

The minimum phonon occupation number for the case of an ideal, monochromatic driving laser has been discussed in several publications \cite{WisonRae2007, Marquardt2007}.  It is limited by the cavity decay rate, $\kappa$, and the mechanical frequency of the movable end mirror, $\Omega$.  Ground state cooling can be achieved when $\kappa \ll \Omega$ and $\Delta \approx -\Omega$. In practice, a more severe limitation arises from the clamping losses associated with the mechanical support of the movable end mirror.  Proposals to reduce or eliminate clamping noise include the optical levitation of the end mirror, see Ref.~\cite{Swati2010}.

In contrast to the preceding discussion, a derivation of the minimum phonon occupation number $\langle n\rangle_{\rm min}$ clearly requires a quantum mechanical description of the mirror motion. It is given by
\begin{equation}
\langle n\rangle_{\rm min} = \frac{1}{2\pi}\int_{-\infty}^\infty d\omega S_N[\omega],
\end{equation}
where the noise spectral density is given by
\begin{eqnarray}
S_N[\omega] &\approx& \int_{-\infty}^\infty dt e^{i\omega t} \langle \langle \hat{a}^\dagger(t)\hat{a}(0)\rangle\rangle_{\rm av} \nonumber \\
&=& \frac{ (2 \gamma+\kappa) \sigma_{\rm opt}(\omega)+\Gamma \sigma_{\rm th}(\omega)}{ |\Lambda(\omega)|^2},
\label{Spectrum}
\end{eqnarray}
see Appendix B, and $\langle\rangle_{\rm av}$ is an average over the classical noise. Here
\begin{eqnarray}
\sigma_{\rm opt}(\omega) &=& \frac{4 g_0^2 |B_0|^2}{(2 \gamma+\kappa)^2+4(\omega+\Delta)^2} \left| \chi_M^{-1}(\omega)\right|^2,\nonumber \\
\sigma_{\rm th}(\omega) &=& n_M \left| \chi_M^{-1}(\omega)+\sigma^*(\omega)\right|^2+(n_M+1) |\sigma(\omega)|^2,
\nonumber \\
\Lambda(\omega)&=& \chi_M^{-1}(\omega) \chi_M^{-1*}(-\omega)-2 i \Omega \sigma(\omega),
\nonumber \\
\sigma(\omega) &=& g_0^2 |B_0|^2 \left[ \chi_R(\omega)-\chi_R^*(-\omega)\right]
\nonumber \\
B_0 &=& \sqrt{P \kappa/ \hbar \omega_c} \chi_R(0),
\end{eqnarray}
and we have introduced the mechanical and optical response functions
$$
\chi_M(\omega) = \frac{1}{\Gamma/2-i(\omega-\Omega)},\,\,\,\,
\chi_R(\omega) = \frac{1}{\kappa/2-i(\omega+\Delta)}.
$$
From the cantilever occupation number spectrum, it is a simple matter to find in the weak coupling limit ($\Gamma_{\rm opt} \ll \kappa$),
\begin{equation}
\langle n\rangle_{\rm min} = - \frac{ (2\gamma+\kappa)(\kappa^2+4(\Delta-\Omega)^2)(\kappa^2+4(\Delta+\Omega)^2)}{16 \Delta \Omega \kappa \left(\left(2 \gamma+\kappa\right)^2+4 \left(\Delta-\Omega\right)^2\right)}.
\end{equation}
This expression reduces to the result of Ref.~\cite{Marquardt2007} for $\gamma = 0$.  In the good cavity limit, $\gamma \ll \kappa \ll \Omega$ and for $\Delta =-\Omega$, $\langle n\rangle_{\rm min}$ becomes
\begin{equation}
\langle n\rangle_{\rm min} =\frac{(2\gamma +\kappa)\kappa}{16 \Omega^2}.
\end{equation}

\begin{figure}[t]
\center
\includegraphics[width= 0.44 \textwidth]{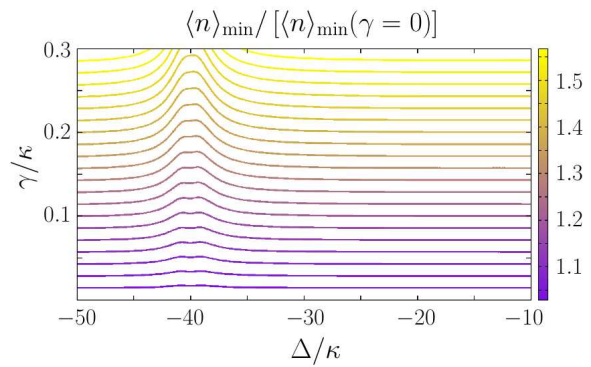}
\label{nmin}
\caption{(Color Online) Minimum occupation number in the absence of mechanical damping compared to the $\gamma=0$ case as a function of $\gamma/\kappa$ and detuning $\Delta/\kappa$.  In this example $g_0 = 50$ rad/s, $\kappa = 2\cdot 10^5$ rad/s,  $\Omega = 40 \kappa$ and $N_{\rm max} = 10^{11}$, where $N_{\rm max} = 4 P /\hbar \omega_c \kappa$ is the maximum number of photons supported by the cavity.}
\end{figure}
For $\Omega = 40 \kappa$ and $\gamma = 0.1\kappa$ this yields a 20 percent increase in the minimum occupation number.  In other words, for a laser with narrow linewidth compared to the cavity decay rate $\kappa$ there is no significant increase in the minimum occupation number and phase noise does not pose a significant problem for ground state cooling, in agreement with the conclusions of Ref.~\cite{Rabl2009}.

In the strong cooling regime, ($g_0^2 \left|B_0\right|^2 \gg \Gamma \kappa$), the effects of a relatively narrow laser linewidth are even less dramatic and likewise do not pose a serious problem for ground state cooling.  This is illustrated in Fig.~6, which shows the minimum occupation number compared to the zero-linewidth case as a function of $\gamma$ and detuning $\Delta$.  We observe that the effects of laser linewidth are suppressed near $\Delta = -\Omega$.  For very large $\gamma/\kappa$, though, the occupation number does increase significantly. 

\section{Parametric driving}

Following an original proposal by Braginsky, Vorontsov and Thorne~\cite{Braginsky1980}, Clerk {\em et al.}~\cite{Clerk2008} have recently shown that by modulating the driving laser frequency at the mechanical frequency $\Omega$ and driving on cavity resonance $\omega_{\rm c}$, 
\begin{equation}
E(t) = \sqrt{\frac{P}{c \epsilon_0 \sigma}} \sin(\Omega t),
\end{equation}
where $P$ is the maximum laser power and $\omega_\ell = \omega_{\rm c}$, it is possible to minimize back-action heating of one of the quadratures of COM mirror motion
\begin{eqnarray}
\hat{X} &=& \frac{1}{\sqrt{2}}\left[\hat{a}e^{i \Omega t}+\hat{a}^\dagger e^{-i \Omega t}\right] \nonumber \\
\hat{Y} &=& -\frac{i}{\sqrt{2}}\left[\hat{a} e^{i\Omega t}-\hat{a}^\dagger e^{-i\Omega t}\right].
\label{quadratures}
\end{eqnarray}
Because laser phase noise places additional limitations on cooling, we expect that the phase noise will also increase the back-action heating of one of the quadratures.

We consider a measurement of one of the quadratures in the weak coupling limit, $g_0^2|B_0|^2\ll \kappa^2$. With these constraints and following a similar method to that outlined in Appendix B, it is possible to find the time averaged variance of the cosine and sine quadratures. For $\gamma,\Gamma \ll \kappa, \Omega$ we find
\begin{eqnarray}
\Delta \hat{X}^2&=& \frac{\Omega}{2 \pi}\int_0^{\frac{2\pi}{\Omega}} \Delta \hat{X}^2(t) dt = \frac{1}{2}\left(2 n_M +1\right)+ \nonumber \\
&+&48 \frac{|b_0|^2 g_0^2 \kappa}{\Gamma}\Big[\frac{\kappa  (\kappa^2+12 \Omega^2)}{(\kappa^2+4\Omega^2)^2(\kappa^2+16 \Omega^2)} \nonumber \\
&+& 3 \gamma\frac{512 \Omega^8+352 \Omega^6 \kappa^2-104 \kappa^4 \Omega^4 -20\kappa^6 \Omega^2-\kappa^8}{(\kappa^2+\Omega^2)(\kappa^2+4\Omega^2)^3(\kappa+16 \Omega^2)^2} \Big]\nonumber \\
\Delta \hat{Y}^2&=&\frac{\Omega}{2 \pi}\int_0^{\frac{2\pi}{\Omega}} \Delta \hat{Y}^2(t) dt = \Delta \hat{X}^2+ \nonumber \\
&+&32 \frac{|b_0|^2 g_0^2 }{\Gamma}\Big[\frac{  (4 \Omega^2 -\kappa^2)}{(\kappa^2+4\Omega^2)^2} \nonumber \\
&-& \gamma\frac{32 \Omega^6+24 \kappa^2 \Omega^4+16 \kappa^4 \Omega^2-3 \kappa^6}{\kappa(\kappa^2+\Omega^2)(\kappa^2+4\Omega^2)^3} \Big],
\label{TimeAvgVariance}
\end{eqnarray}
where $|b_0|^2 = P/ \hbar \omega_c$, and we have taken a time average. In the good cavity limit $\kappa \ll \Omega$ these expressions reduce to
\begin{eqnarray}
\Delta \hat{X}^2 &\approx&  \frac{1}{2}(n_M+1)+g_0^2 |b_0|^2 \frac{9 \kappa (\kappa+2 \gamma)}{4\Gamma \Omega^4},\nonumber \\
\Delta \hat{Y}^2 &\approx&\frac{1}{2}(n_M+1)+g_0^2 |b_0|^2 \frac{9 \kappa (\kappa+2 \gamma)}{4\Gamma \Omega^4}\nonumber \\
&+& 8 g_0^2 |b_0|^2 \frac{\kappa-2 \gamma}{\kappa \Gamma\Omega^2}+16 g_0^2 |b_0|^2 \frac{\gamma \kappa}{\Gamma \Omega^4}.
\label{quad}
\end{eqnarray}

Equations~(\ref{TimeAvgVariance}) and (\ref{quad}) show that the laser phase noise results in an increase in fluctuations of the quadratures of COM motion. What may appear surprising is that a contribution proportional to $\gamma$ is preceded by a minus sign in $\Delta {\hat Y}^2$. Keeping in mind that these results are only valid in the limit $\gamma, \Gamma \ll \kappa, \Omega$, we emphasize that this {\em does not} imply that phase diffusion results in a reduction in fluctuations in the cosine quadrature, but merely that its variance increases more slowly than the variance of the sine quadrature. This can be understood intuitively from the fact that while a perfectly sinusoidal driving field provides an optimal back-action evasion method for the cosine quadrature~\cite{Clerk2008}, phase noise in the driving laser translates into intracavity intensity fluctuations about zero frequency. These fluctuations increasingly overwhelm the back-action evasion provided by the sinusoidal drive, resulting in the effect of the sinusoidal drive being reduced in relative importance, and additional heating in each quadrature due to laser phase noise. In the limit $\gamma \gg \kappa$ one would expect both quadratures to be heated equally, which means the variance of the sine quadrature must `catch up'  with that of the cosine quadrature.

\begin{figure}[t]
\center
\includegraphics[width= 0.48 \textwidth]{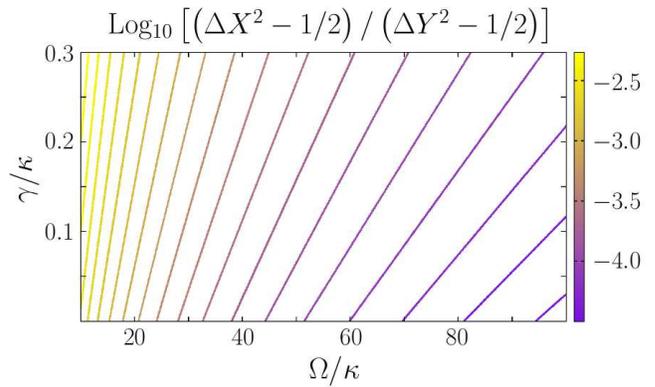}
\label{parametricdrive}
\caption{(Color Online) Contour plot of the heating of the cosine quadrature due to the drive laser compared to that of the sine quadrature as a function of mechanical frequency $\Omega$ and laser linewidth $\gamma$.  In this plot we have assumed $n_M = 0$.  Note the increase in the cosine quadrature heating due to the laser linewidth.}
\label{Parametric}
\end{figure}

Because the back-action heating of the sine quadrature is proportional to the mean intra-cavity photon number, the effect of back-action can easily be limited to an acceptable level, even in the presence of laser phase noise. A comparison of the heating of the cosine quadrature to the sine quadrature due to phase diffusion is shown in Fig.~\ref{Parametric} as a function of mechanical frequency $\Omega$ and laser linewidth $\gamma$. We see that the cosine quadrature is heated by nearly a full order of magnitude for a laser linewidth of $\gamma = 0.3 \kappa$.  This implies that larger laser linewidths can hinder this back-action evasion method and well stabilized lasers are necessary for employing this method.

\section{Conclusions}

We have analyzed the effects of laser phase noise in the dynamics of a generic optomechanical system, considering both single-frequency driving that can result either in back-action damping or mechanical amplification, and parametric driving useful for the generation of squeezing and back-action evading measurement schemes. We showed that laser phase noise reduces the effectiveness of backaction damping and softens the effects of a mechanical frequency shift.  Additionally, we observed an increase in the minimum phononic occupation number of the mechanical element that remains however modest for $\gamma \ll \kappa$.  It was concluded that ground state cooling can easily be achieved with a well stabilized laser. When extending the results of Clerk {\it et. al.}~\cite{Clerk2008} on back-action evasion to include the influence of laser phase noise we showed that the laser phase noise results in additional heating of both sine and cosine quadratures, as expected.  Overall, though, we have shown that for narrow laser linewidths such that $\gamma \ll \kappa$ the contribution from this noise source remains small. Future work will extend theses results to the preparation and detection of nonclassical mechanical states, including squeezed states, number states, and Schr{\"o}dinger cat states, and the analysis of quantum state transfer between mechanical and electromagnetic degrees of freedom. 

\begin{acknowledgments}
 We thank S. Singh, D. Goldbaum, S. Steinke and E. M. Wright for stimulating discussions, and P. Rabl for insightful comments on radiation pressure induced cooling rates.  This work is supported by the US National Science Foundation, the DARPA ORCHID program, the US Army Research Office, US Office of Naval Research, and the University of Arizona/NASA Space Grant.
\end{acknowledgments}

\appendix
\section{Cold damping and mechanical frequency shift}

The dynamics of the mirror and the intracavity light field are given by Eqs.~(\ref{classical_langevin}). We consider in the following the simplified case of classical COM motion at frequency $\Omega$ in the absence of light field,
\begin{equation}
\label{ansatz}
x(t) = x_0 \sin(\Omega t).
\end{equation}
This simplification is sufficient to determine the cooling rates and mechanical frequency shifts from an initially classical state. (Note that $x_0$ is bounded from below by the zero point motion $x_0 \geq \sqrt{\hbar/2 M \Omega}$. )

We proceed by substituting the ansatz~(\ref{ansatz}) into the Eq.~(\ref{classical_langevin}) for the light field and solve for $\beta(t)$. Integrating that equation formally gives
\begin{equation}
\beta\left(t\right) = \beta\left(0\right) e^{\left[i\Delta-\frac{1}{2}\kappa\right]t-i\epsilon [\cos(\Omega t)-1]}+\beta_P(t),
\end{equation}
where $\epsilon = \omega_c x_0/ L \Omega$ and $\beta_P(t)$ is the contribution of the driving laser field, given explicitly by
\begin{widetext}
\begin{equation}
\beta_P(t) = \sqrt{\kappa P} e^{[i\Delta-\frac12\kappa]t-i \frac{\omega_c x_0}{\Omega L}cos(\Omega t)} \sum_{n = -\infty}^{\infty} i^n J_n(\epsilon)\int_0^t e^{i (n \Omega -\Delta ) s +\frac{1}{2}\kappa s}e^{i\phi (s)} ds.
\end{equation}
\end{widetext}
In deriving this expression we have used the Jacobi-Anger expansion on the $\exp[i \epsilon \cos(\Omega s)]$ term, and $J_n(z)$ is a Bessel function of the first kind.

In the following we ignore the free transients compared to the relevant driven contribution to the intracavity field, resulting in the intracavity normalized intensity
\begin{widetext}
\begin{equation}
|\beta_P(t)|^2 = \kappa P e^{-\kappa t} \sum_{n_1,n_2 = -\infty}^{\infty} i^{n_1-n_2} J_{n_1}(\epsilon)J_{n_2}(\epsilon) \int_0^t \int_0^t e^{i (n_1\Omega-\Delta)s-i (n_2 \Omega-\Delta)s'+\frac{1}{2}\kappa (s+s')} e^{i (\phi(s)-\phi(s'))}ds'ds.
\end{equation}
\end{widetext}
We include the effect of the Lorentzian spectrum of the driving laser via an ensemble average over the random phase noise~(\ref{noise}),
\begin{equation}
\left\langle e^{i [\phi(s)-\phi(s')]}\right\rangle _{\rm av} = e^{-\gamma \left|s-s'\right|}
\end{equation}
to find the ensemble-averaged intracavity normalized intensity $\langle |\beta_P(t)|^2\rangle_{\rm av}$. In most cases of practical interest in cavity optomechanics we have that $\epsilon \ll 1$. Keeping then linear terms in $\epsilon$ only we find
\begin{equation}
\left\langle |\beta_P(t)|^2\right\rangle_{\rm av} \approx M L [ |\beta_0|^2- \Omega_{\rm opt}^2 x(t)-\Gamma_{opt} \dot{x}(t)],
\end{equation}
where $|\beta_0|^2$ is given explicitly in Eq~(\ref{beta0}).  Considering $|\Omega_{\rm opt}|^2 \ll \Omega^2$ for our ansatz, we have the effective mechanical frequency given by
\begin{equation}
\Omega_{\rm eff} = \sqrt{\Omega^2+\Omega_{\rm opt}^2} \approx \Omega+\frac{1}{2\Omega}\Omega_{\rm opt}^2 = \Omega+\Delta\Omega_{\rm opt}
\end{equation}

The explicit form of the frequency shift $\Delta \Omega_{\rm opt}$ is given in Eq.~(\ref{Gamma opt}). This shift is due to the component of the light field that is in-phase with the mirror oscillations. The mechanical damping $\Gamma_{\rm opt}$ is given explicitly in Eq.~(\ref{omega}), and is due to the out-of-phase components of the light field.

\section{Cantilever occupation number spectrum}

Our starting point is the linearized equations of motion~Eqs.~(\ref{LinearA},\ref{LinearD},\ref{LinearB}) and the electric field $|E|^2 = P/(c \epsilon_0 \sigma)$.  These equations of motion are conveniently manipulated in the Fourier domain. Introducing the Fourier transform of an arbitrary operator $\hat{c}(t)$ as
\begin{align}
C[\omega] &= \int_{-\infty}^\infty dt \hat{c}(t) e^{i\omega t}\\
C^\dagger[\omega] &= \int_{-\infty}^\infty dt \hat{c}^\dagger(t) e^{i\omega t},
\end{align}
which leads to the Fourier space coupled equations for the cantilever and light field
\begin{equation}
A[\omega] = \chi_M(\omega)\left[-\sqrt{\Gamma} A_{in}[\omega]+i \frac{g_{\rm 0}}{2 \pi} F[\omega]\right],
\nonumber
\end{equation}
\begin{equation}
D[\omega] = \chi_R(\omega)\left[-\sqrt{\kappa} D_{\rm in}[\omega]+i \frac{g_{\rm 0}}{2 \pi} \bar{B}[\omega]\star\left(A^\dagger[\omega]+A[\omega]\right)\right],
\nonumber
\end{equation}
where
\begin{widetext}
\begin{eqnarray}
F[\omega] &=& -\sqrt{\kappa}\Big[\bar{B}^\dagger[\omega]\star \left(\chi_R[\omega] D_{\rm in}[\omega]\right) +
+\bar{B}[\omega]\star(\chi_R^*(-\omega) D^\dagger_{\rm in}[\omega])\Big]
+ i\frac{g_{\rm 0}}{2 \pi}\Big\{ \bar{B}^\dagger[\omega]\star\left[\chi_R(\omega) \left(\bar{B}[\omega]\star(A[\omega]+A^\dagger[\omega])\right)\right]\nonumber \\
&-&\bar{B}[\omega]\star\left[\chi_R^*(-\omega) \left(\bar{B}^\dagger[\omega]\star(A[\omega]+A^\dagger[\omega])\right)\right]\Big\}.
\nonumber
\end{eqnarray}
Here the convolution of two arbitrary functions is as usual
\begin{equation}
f(\omega)\star g(\omega) \equiv \int_{-\infty}^\infty dx f(x) g(\omega-x),
\nonumber
\end{equation}
and for small laser linewidths $\gamma \ll \kappa$ we have
\begin{equation}
\bar{B}[\omega] \approx -2 \pi \chi_R[\omega]\sqrt{\kappa} b_0 \delta(\omega).
\nonumber
\end{equation}
With this explicit form, $F(\omega)$ reduces to
\begin{equation}
F[\omega] \approx -\sqrt{\kappa}\Big[\bar{B}^\dagger[\omega]\star \left(\chi_R[\omega] D_{\rm in}[\omega]\right)+
+\bar{B}[\omega]\star(\chi_R^*(-\omega) D^\dagger_{\rm in}[\omega])\Big]\nonumber
+ i 2 \pi g_{\rm 0}\left|\bar{B_0}\right|^2\left( \chi_R(\omega) -\chi_R^*(-\omega)\right) (A[\omega]+A^\dagger[\omega]) ,
\nonumber
\end{equation}
where $\bar{B}_0 = \sqrt{\kappa} b_0 \chi_R[0]$.  In this approximation we can easily find a solution for $A[\omega]$ as
\begin{equation}
A[\omega] \approx \frac{\chi_M(\omega)}{\Sigma[\omega]} \Big[ i g_0 F_0[\omega]-\sqrt{\Gamma} A_{\rm in}[\omega]
+\sqrt{\Gamma} g_0^2 |B_0|^2 \chi_M^*(-\omega) \left[A_{\rm in}[\omega]+A_{\rm in}^\dagger[\omega] \right]\times \left\{\chi_R(\omega)-\chi_R^*(-\omega)\right\}\Big],
\end{equation}
where
\begin{equation}
\Sigma[\omega] = 1+g_0^2 |B_0|^2 \left(\chi_M(\omega)-\chi_M^*(-\omega)\right)\left(\chi_R(\omega)-\chi_R^*(-\omega)\right),
\nonumber
\end{equation}
and
\begin{equation}
F_0[\omega] = -\sqrt{\kappa}\Big[\bar{B}^\dagger[\omega]\star \left(\chi_R(\omega) D_{\rm in}[\omega]\right) +
\bar{B}[\omega]\star(\chi_R^*(-\omega) D^\dagger_{\rm in}[\omega])\Big].
\nonumber
\end{equation}
\end{widetext}
With the two-frequency noise input correlations:
\begin{eqnarray}
\langle D_{\rm in}^\dagger (\omega)D_{\rm in}(\omega')\rangle &=& 0\nonumber \\
\langle D_{\rm in} (\omega) D_{\rm in}^\dagger(\omega')\rangle &=& 2\pi\delta(\omega+\omega')\nonumber \\
\langle A_{\rm in}^\dagger (\omega) A_{\rm in}(\omega')\rangle &=& 2 \pi n_{M}\delta(\omega+\omega')\nonumber \\
\langle A_{\rm in}(\omega) A_{\rm in}^\dagger(\omega')\rangle &=& 2\pi(n_{M}+1)\delta(\omega+\omega'),\nonumber
\end{eqnarray}
which are equivalent to the two-time correlations of Eqs.~(\ref{corr_din}, \ref{corr_ain}), it is a simple matter to find 
\begin{equation}
S_N[\omega] = \int_{-\infty}^\infty \frac{d\omega'}{2\pi} \langle\langle A^\dagger[\omega]A[\omega']\rangle\rangle_{\rm av}.
\label{averageN}
\end{equation}




%

\end{document}